# TH*: Scalable Distributed Trie Hashing

ARIDJ MOHAMED [1,2], ZEGOUR DJAMEL EDINNE [1]

[1] Institut National d'Informatique
Alger, Alegria

[2] University Hassibe benbouli  Department, University Name, Company
Chlef, Algeria

**Abstract**

In today's world of computers, dealing with huge amounts of data is not unusual. The need to distribute this data in order to increase its availability and increase the performance of accessing it is more urgent than ever. For these reasons it is necessary to develop scalable distributed data structures.
In this paper we propose a TH* distributed variant of the Trie Hashing data structure. First we propose Thsw new version of TH without node Nil in digital tree (trie), then this version will be adapted to multicomputer environment. The simulation results reveal that TH* is scalable in the sense that it grows gracefully, one bucket at a time, to a large number of servers, also TH* offers a good storage space utilization  and  high query efficiency special for ordering operations.

## 1. Introduction

A multi-computer consists of set of workstations and PCs interconnected by a high speed network such as the Ethernet and TMT. It is well known that multi-computers offer best price-performance ratio; thus offering some new perspectives to high performances applications [11, 13].

In order to achieve these performances, a new class of data structures has been proposed. It is called Scalable Distributed Data Structures (SDDS) [3,4,5,9,8,10,12, 14,16] and is based on the client/server architecture. This new structure supports the parallel processing that does not require the central processing of the addresses. Data is typically stored in the distributed main memory (DRAM). An SDDS may easily handle very large files and their access is achieved in a fraction of the disk access time. An SDDS propagates to new sites through splitting when sites are saturated. The splitting process is transparent to the applications.

All SDDSs support key searches where some supports range and/or multi key searches. However, every client has its own file picture. The updating of the file structure are not sent to the clients in a synchronous manner. A client can make addressing error when using its own file picture. Each server verifies the address of the received request and is routed to another server if an address error is detected. The server that processes the requests then sends an adjusted message to the client that made the addressing error. This message is called Image Adjustment Message (IAM).The IAM allows the client to adjust its image to avoid making the same error again. However, its image is not necessarily globally exact.

**Related work**

LH*[8] is the first proposed SDDS.  It is based on the linear hashing technique [17]. It achieves good performance for single-key operations, but range searches are not performed efficiently. To rectify this, order preserving structures was proposed. Among order preserving SDDSs, we recall RP* family [18],BDST[19] and ADST[20].

Rp*, based on the B+-tree technique and BDST, based on balanced binary search tree achieve good performances for range searches and a reasonably low ,but not constant worst-case cost for single key operations. ADST (Aggregation in Distributed Search Tree) obtaining a constant single-key query cost, like LH*, and an optimal cost for range queries, like RP* and DRT* but a logarithmic cost for insert queries producing a split.

All these techniques (except RP*N) use an index witch is stored at the servers and/or clients RAMs. This index may take up large  space in the RAM.

In this paper, we propose a new SDDS, called   TH*, that adapts the TH to distributed environments. The remainder of the paper is organized as follows.  In section 2 we review Trie Hashing [6] and in Section 3 a new TH version which eliminates nil nodes is discussed. The principles and algorithms of the new SDDS TH* are introduced in Section 4 and section 5 discusses the results of an evaluation of TH* using simulation. We conclude and summarize the paper in Section 6.

## 2. Trie hashing

Trie hashing is one of the most powerful access methods used for dynamic and ordered files. The TH file is a set of records identified by primary keys. Keys consist of a finite ordered sequence of characters. Records are stored in buckets and each bucket may contain a fixed or a variable number of records (Fig.1.b). The file is addressed thought the consisting of the nodes where each node is either a leaf or an internal.(Fig.1.a).

An internal node contains a pair of values (d,i) where d is a character and i is a number representing the position of the character d in the key. However the leaf node contains the bucket address or the value Nil indicating that no buckets are associated with the leaf.

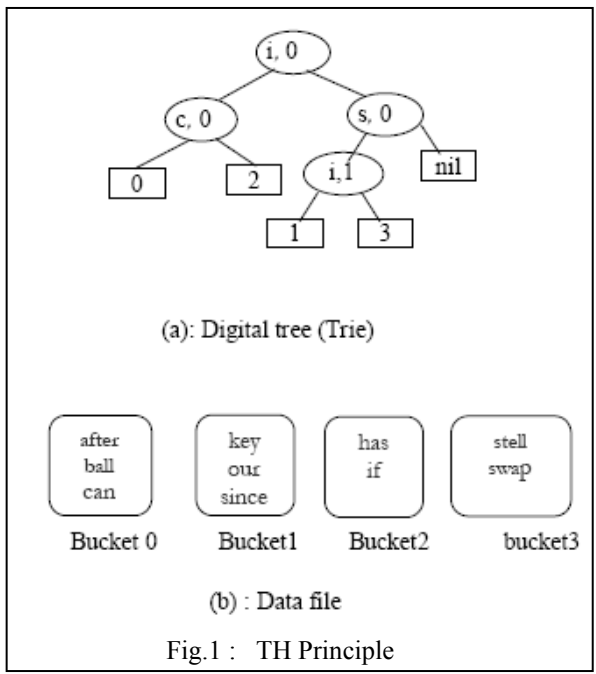

Fig.1 : TH Principle

The insertion operation may cause the expansion of the file and the trie; however the deletion operation may cause a contraction. All the algorithms related to TH are taken from [6]. The load factor of the file is about 70% for random insertions and of 60% to 70% for ascending insertions. The key search takes at most one disk access and insertion takes 2 or 3 disk access. The behaviour and performance analysis of TH can be found in[1,2,7,15].

## 3. THwn: trie hashing without Nil nodes

Given that in the trie the Nil nodes appear when more than one digit is necessary for the split string, Litwin's method solves the problem as follows:

Let $C="d_0d_1d_2......d_n"$ be the split string necessary to split a bucket m into the bucket m' the resulting trie will have the form described in Fig 2.

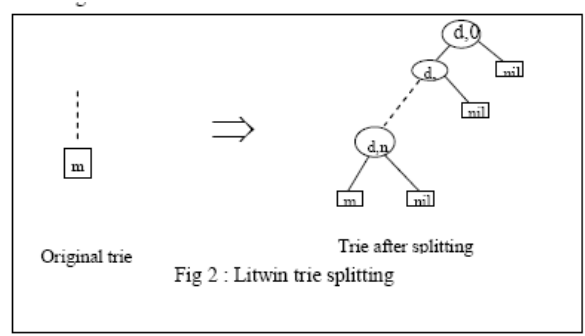

Fig 2 : Litwin trie splitting

3.1 Proposition:

To eliminate the nil nodes we propose that they will be replaced by the address of the new allocated bucket (m').In this case the trie will have the following form given in Fig 3.

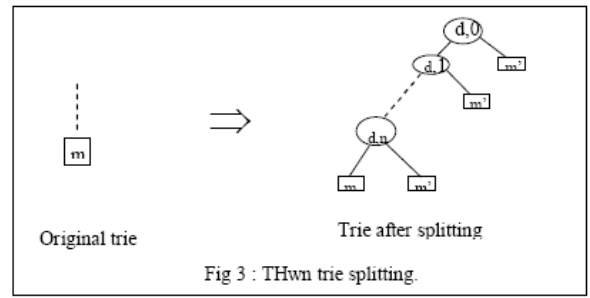

Fig 3 : THwn trie splitting.

This proposition gives a new trie hashing scheme we named THwn and which stands for "Trie Hashing Without Nil nodes."

3.2 Example :

To illustrate the principle of THwn we insert the following keys: **"abmf", "abnm", "acnm", "aczm", "aczh"** in the file and we assume that the capacity of a bucket is equal to 4. The insertion of the **"abmf", "abnm", "acnm", "aczm",** is done in bucket 0, the insertion of the **"aczh"** causes a collision; in this case the splitting string is "acn". The trie after splitting is show in Fig.4.

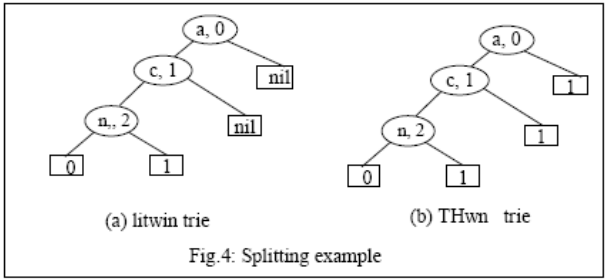

Fig.4: Splitting example

Our proposal is justified by the behavioural analysis of the bucket range. By observing the example given in Fig4, we note that initially all the keys will be inserted in the bucket because the range of this bucket is [ _, | ] (where _ represents the smallest digit and | the greatest digit). After the proposed splitting we obtain the configuration given in Fig 5.b. However the application of Litwin's algorithm gives the configuration shown Fig 5.a.

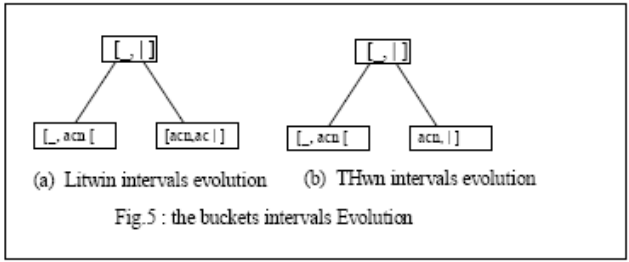

Fig.5 : the buckets intervals Evolution

### 3.3 The THwn Algorithms

The algorithm for searching and inserting remain similar to those proposed in [6]. However for bucket splitting we propose the following algorithm given in Fig6 and its illustration using an example is given in Fig 7.

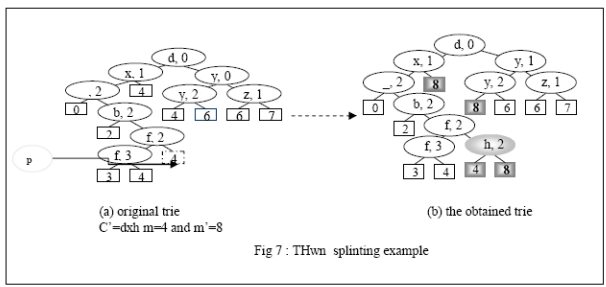

Fig 7 : THwn splinting example

## 4. Distributed trie hashing (TH *)

In this section we propose the extension of THwn to a distributed environment, the obtained schema is named TH*

### 4.1 Concepts

As previously mentioned, TH* is based on the client/server architecture. Each client contains a partial trie which represents the client image of the distributed file and is being updated gradually until we obtain the real trie. Any client can enter the system with an empty trie and each server has a Bucket containing the records of the file, the trie and an interval [Min, Max ] where Min is the smallest value the server can contain and Max the largest value.

Initially the system contains only the server 0 with an empty bucket. Its interval is ]"_......","|…….."] as previously introduced and with an empty trie (leaf with value 0). The file expands by splitting due to collisions.

At each collision there is a distribution of the keys (splitting of overflowed server) to another server. The number of servers is conceptually infinite and each server can be determined in a static or dynamic way. When an addressing error occurs, part of the server trie is transferred to the client for updating its trie

### 4.2 Example of TH* File

To illustrate our method we consider a system composed of 4 clients and several servers. We suppose that the server's capacity is 4. The following list of insertions is represented by a set of pairs (Client_Number, keys).
(1 js), (1 hw), (3, c), (2, gwmr), (3, g), (2, km), (4, zur), (1,ewg), (3, lewhv), (2, nrq), (3, mf), (4, pem), (4, rl), (2, bqyg), (3, v), (1, j), (2, qcm), (4, czxav), (2, lhgd), (3, z), (1, lrz), (3, kiyfg), (4, pbtpr), (3, hpqtp), (4, h).
Fig 8 shows the final state of the system after 25 insertions

| Client1 | Client2 | Client3 | Client4 |
|---|---|---|---|
| (trie) | (trie) | (trie) | (trie) |

| | Server 0 | Server 1 | Server 2 | Server 3 |
|---|---|---|---|---|
| Trie | Sup ='e' | Sup ='h' | Sup ='l' | Sup ='r' |
| Bucket | bgvg<br>c<br>cav<br>cwg | h<br>hpgtp<br>hw | Lewhv<br>lhgd<br>lrz | pem<br>rl<br>qcm<br>pbtpr |

| | Server 4 | Server 5 | Server 6 | Server 7 | Server 8 |
|---|---|---|---|---|---|
| Trie | 4 | 5 | 6 | 7 | 8 |
| | Sup='g' | Sup='|' | Sup='n' | Sup='k' | Sup='j' |
| Bucket | g<br>gwmr | v<br>z<br>zur | mf<br>nrq | kiyfg<br>km | j<br>is |

Fig.8 : TH* file example.

## 4.3 TH* algorithms

### 4.3.1 addressing :
Transformation key --> server address

To realise an operation on a TH* file, the client calculates the address of a suitable server by traversing its trie which represents its image on the TH * file. When the server receives a client request, it checks if the key is in its interval. Two cases can occur:

**Case 1** The key is in the server interval: the server treats the request and sends back an Image Adjustment Message (IAM) to the client if an addressing error had being detected. The client uses the IAM to update its image.

**Case2** There is an addressing error: the server recalculates the address using its cached file image (trie) and forwards the message to another server as illustrated in Fig 9.

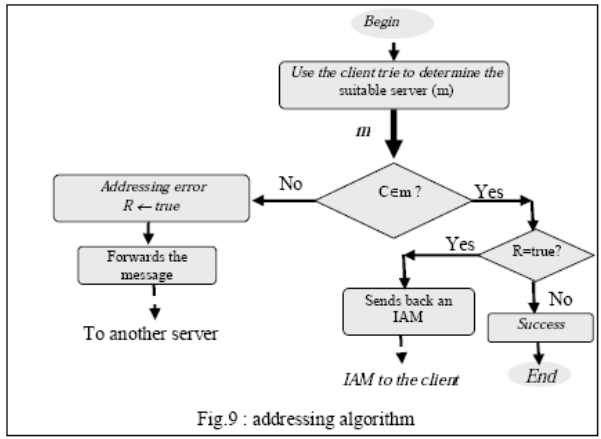

Fig.9 : addressing algorithm

### 4.3.2 Adjusting client image

In the case of an addressing error by the client, the last server participating in the forwarding process sends back the adjustment image message to the client. The client then updates its trie using the algorithm given in Fig 10 which is illustrated by the examples given in Fig 11 and Fig 12.

**Step 1 : constructing the IAM at server using the following algorithm:**
in this step the server determines the sub trie to send to the client in order to correct its false image (trie). This algorithm can be as follows:
Let $Cm_{client}$ : be the maximum key attributed to the concerning server by the client trie.(sent in the client query)
1. [determination of server maximum key]
   Determine the $Cm_{server}$ the maximum server key by traversing the local trie.
2. [determination of common prefix]
   let prefix=$Cm_{server}$ - $Cm_{client}$
3. [construction of IAM]
   if prefix=$Cm_{server}$ then    IAM= server trie
                     Else    IAM= the sub trie
of server giving the prefix string
4. [send IAM (sub trie) to the client]

**Step 2: updating client trie:**
1. [receive the IAM from server]
   Let S ← the sub trie contained in the IAM
2. [trie updating]
   Let  P  be the pointer of the leaf contains m ( resulting from search algorithm)
   Let m' be the server address in   the latest leaf of S
   Replace the leaf m by the S (attach S to the trie)
    While forward(p) = m do
   Replace m by m'
   P← forward(p).

Fig 10 : Adjusting client image algorithm

**Examples:**

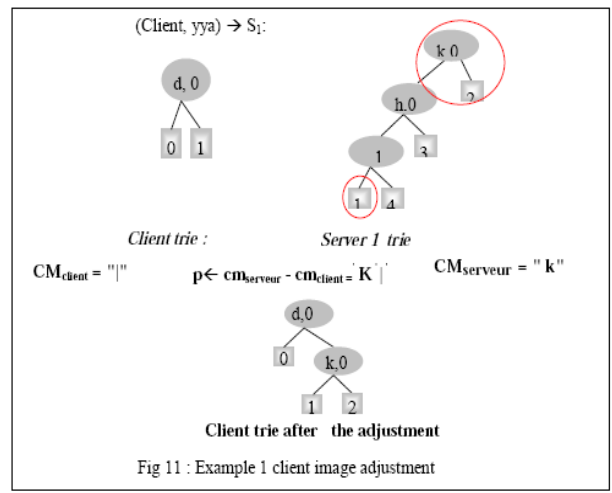

Fig 11 : Example 1 client image adjustment

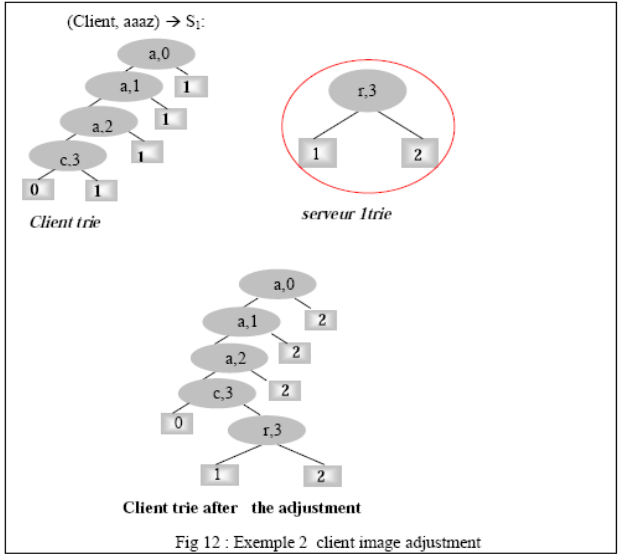

Fig 12 : Exemple 2 client image adjustment

### 4.3.3 Insertion

The insertion requires in a first instance the check if the key exists using the search algorithm. If the key does not exist in the file, it will be inserted in the appropriate server. However, if this server is full the splitting algorithm is then used. Fig. 13 summarizes the principle of an insertion operation in a SDDS TH*

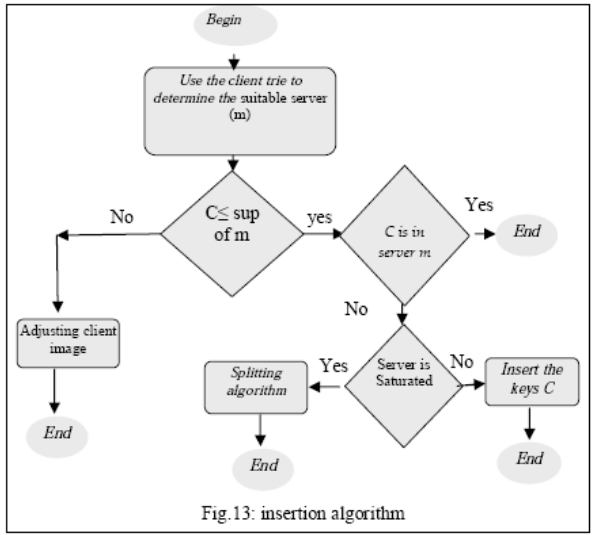

Fig.13: insertion algorithm

### 4.3.3 Server splitting

The splitting algorithm is called by the overflowing server. It allocates a new server and redistributes the keys. The algorithm is given in Fig14

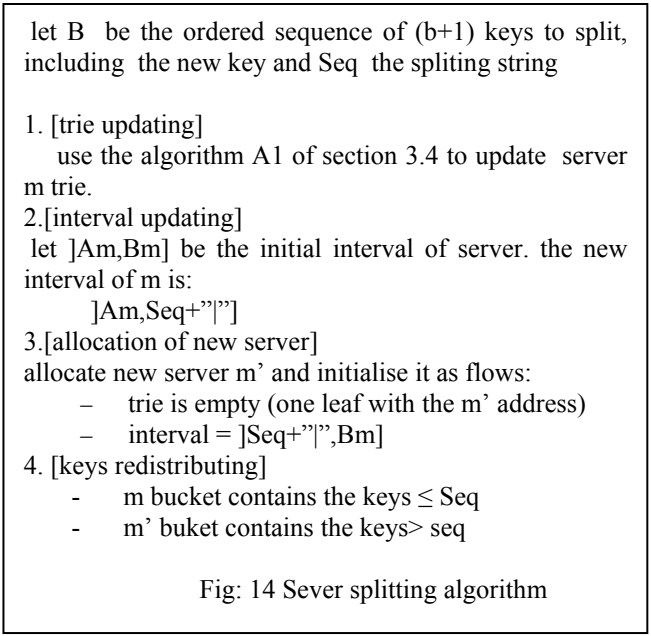

let B be the ordered sequence of (b+1) keys to split, including the new key and Seq the spliting string

1. [trie updating]
   use the algorithm A1 of section 3.4 to update server m trie.
2. [interval updating]
   let $]A_m, B_m]$ be the initial interval of server. the new interval of m is:
       $]A_m, Seq+"|"]$
3. [allocation of new server]
allocate new server m' and initialise it as flows:
   – trie is empty (one leaf with the m' address)
   – interval = $]Seq+"|", B_m]$
4. [keys redistributing]
   - m bucket contains the keys $\leq$ Seq
   - m' buket contains the keys> seq

Fig: 14 Sever splitting algorithm

### 4.3.4 range query

The range query operation consists on listing all the keys in a given interval. Since the TH preserves the keys' order (the trie keeps the order of the keys), this operation is be easy to perform. The general algorithm of this operation is realised in two steps. First the client determines all the servers that could contain the keys, and then the servers process the request. The general algorithm of this operation is as follows :( Fig 15.a and Fig 15.b)

**At level of client**

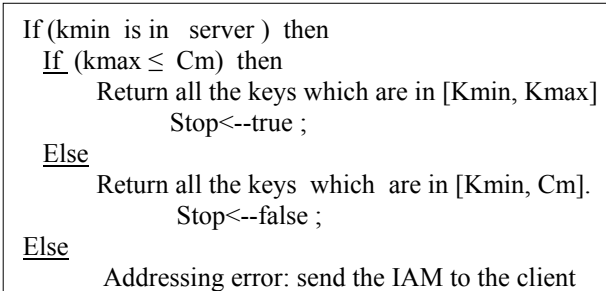

If (kmin is in server ) then
  If (kmax $\leq$ Cm) then
      Return all the keys which are in [Kmin, Kmax]
          Stop<--true ;
  Else
      Return all the keys which are in [Kmin, Cm].
          Stop<--false ;
Else
      Addressing error: send the IAM to the client

Fig 15.a : range query algorithm at level of client

**At level of server**

let [kmin,kmax] be the range query interval

1. Search kmin in the local trie :
   let m be the result, and Cm its maximum keys of m.
2. Send request to the server m

3. at the reception of servers reply one of they tree case is executed :
   Case1 : error address : use the adjustment algorithm to updating the client trie
   Case 2 : Stop=true : server continuing Kmax is attended processes is stooped
   Case 3 : Stop=false : use the server trie to find the next server .

Fig 15.b : range query algorithm at level of server

## 5. Simulation and test results:

5.1 Development environment

We have implemented the SDDSs TH* on a multi-computer composed of 4PCs executing the LINUX system (Mandrake 8) and connected by a high seeped network. The PCs are Pentium V 900 MHz with 128 Mo of RAM. Every machine can be client and/or server

5.2 Some results

**a) Load factor**
We have inserted 1,000,000 keys in a TH* files with different server capacity
(b=50,b=100,b=500,b=1000).
The load factor variations can be summarized in the following graph (Fig 16)
This experiment showed that the server capacity does not affect the behavior of load factor which is giving by figure (Fig.16)..

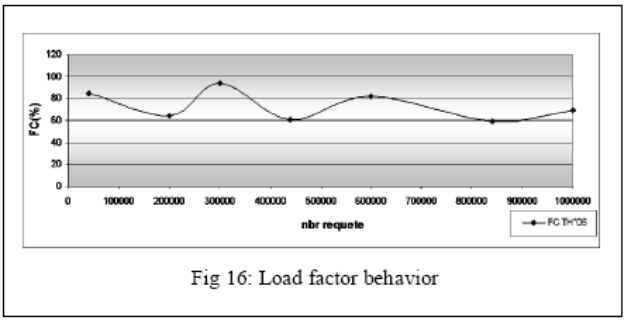

Fig 16: Load factor behavior

**Comment:**
Observation of the figure (Fig 16) shows that:
- TH* offers good storage space utilization, witch it is about (65 % to 93%).
- The server capacity does not influence the behavior of load factor.

**b) Variation of number of servers:**
Under the same hypotheses, we have noticed variations of the servers splitting (Fig 17).

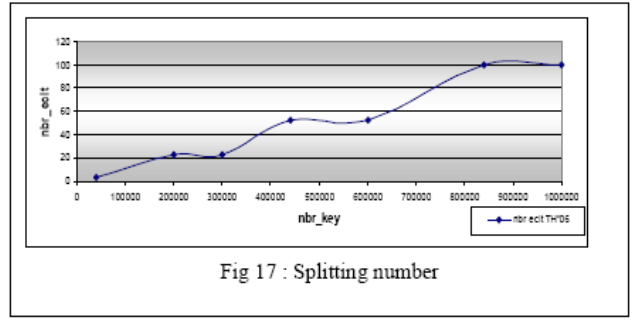

Fig 17 : Splitting number

**Comment:**
The figure (Fig 17) shows that the number of the servers splitting increases linearly with the number of the inserted keys

**c) variation of search and insertion average time**
The study of the two parameters is very important because it makes the method either scalable or not.
The curve of figure (fig.18) represents a research and insertion average time.

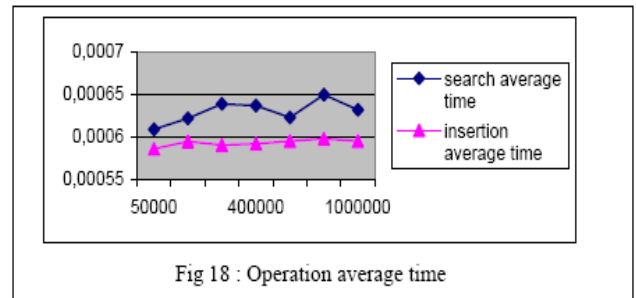

Fig 18 : Operation average time

**Comment:**
The two curves are practically linear. This implies that the research and insertion average time does not increase with the number of keys, which makes the method very scalable.

## 1. Conclusion

This paper presented an adaptation of Trie hashing to distributed environments. The method is based on the client/server architecture and respects the SDDS properties.
The obtained schema is called TH* and its behavioural analysis shows that it is efficient and scalable during insertions and retrievals, with performance close to optimal. It also offers good storage space utilization. Since TH* is an efficient, scalable, distributed data structure, it provides a new method to be used in applications such as next generation databases, distributed data mining, distributed data warehouse, distributed dictionaries, bulletin boards, etc

**Acknowledgments**

The authors would like to thank Litwin Witold and the members of CERIA research group for taking the time discuss the ideas presented here.

**Aridj Mohamed** is assistant professor at university Hassiba Benboali chlef Algeria since 2001.

His area of interest include Software Engineering, distributed System, Databases , access methods and hashing

**Zegour djamel edinne** is professor at institute national d'informatique Algeria.

His area of interest includes Software Engineering, System Analysis and Design, Databases, distributed systems, access methods and Object Oriented Technologies.